\documentclass{osa-article}
\usepackage{subcaption}
\usepackage{float}
\journal{oe}
\articletype{Research Article}
\begin{document}

\title{Temporal mode transformations by sequential time and frequency phase modulation for applications in quantum information science}
\author{James Ashby,\authormark{1,*}, Val\'{e}rian Thiel,\authormark{1,2} Markus Allgaier, \authormark{1} Peru d'Ornellas,\authormark{2} Alex O. C. Davis,\authormark{2,3} and Brian J. Smith\authormark{1,2}}

\address{
\authormark{1}Department of Physics and Oregon Center for Optical, Molecular, and Quantum Science, University of Oregon, Eugene, Oregon 97403, USA\\
\authormark{2}Clarendon Laboratory, University of Oxford, Parks Road, Oxford, OX1 3PU, UK\\
\authormark{3}Laboratoire Kastler-Brossel, UPMC-Sorbonne Universit\'{e}s, CNRS, ENS-PSL  Research University, Coll\'{e}ge de France, 75005 Paris, France
}

\email{\authormark{*}oqt@uoregon.edu} %% email address is required

% \homepage{http:...} %% author's URL, if desired

%%%%%%%%%%%%%%%%%%% abstract %%%%%%%%%%%%%%%%
%% [use \begin{abstract*}...\end{abstract*} if exempt from copyright]

\begin{abstract}
Controlling the temporal mode shape of quantum light pulses has wide ranging application to quantum information science and technology. Techniques have been developed to control the bandwidth, allow shifting in the time and frequency domains, and perform mode-selective beam-splitter-like transformations. However, there is no present scheme to perform targeted multimode unitary transformations on temporal modes. Here we present a practical approach to realize general transformations for temporal modes. We show theoretically that any unitary transformation on temporal modes can be performed using a series of phase operations in the time and frequency domains. Numerical simulations show that several key transformations on temporal modes can be performed with greater than 95\% fidelity using experimentally feasible specifications.
\end{abstract}

%%%%%%%%%%%%%%%%%%%%%%%%%%  body  %%%%%%%%%%%%%%%%%%%%%%%%%%
\section{Introduction}

The ability to manipulate the mode structure of light enables numerous technologies, both quantum and classical, ranging from long-distance telecommunications to microscopy. Quantum applications such as quantum key distribution \cite{bb84}, boson sampling \cite{aaronson13}, linear optics quantum computing \cite{kok07} and continuous-variable quantum information processing \cite{weedbrook12} all rely upon unitary transformation of optical modes for the encoding, manipulation and measurement of quantum information. Over the past decade significant progress has been made to implement multimode transformations with high fidelity and low loss by utilizing integrated optical platforms. This work has primarily focused on polarization and spatial degrees of freedom \cite{wang20}. More recently, time and frequency encoding has come to the fore as a promising platform for encoding quantum information in optical fields \cite{humphreys13, humphreys14, lukens17}. When considering a finite temporal and spectral range, temporal modes, analogous to transverse spatial modes such as the Laguerre-Gauss modes that carry orbital angular momentum, form a useful basis in which to encode quantum information and arise naturally in many common quantum light sources such as spontaneous parametric down conversion \cite{nunn13, brecht15}. 

To address the temporal modes of light for quantum applications most efforts have focused on nonlinear optical means for control \cite{Eckstein, reddy2, humphreys14, sinclair14, ansari18}.
The quantum pulse gate (QPG) \cite{Eckstein,reddy2,ansari18} is a key building block for nonlinear optical methods to control temporal modes which acts, in theory, as a temporal mode beam splitter. To realize multimode transformations using the QPG requires sequential application of multiple QPGs with both high selectivity and high efficiency and appropriate relative phase shifting between the ancilla and fundamental modes of the pulse gate \cite{brecht15}, which has yet to be demonstrated with quantum light. Alternatively, linear-optical methods using acousto- and electro-optic temporal phase modulation and dispersive spectral phase manipulation of pulsed modes have shown promise for quantum applications \cite{fan16, wright17, karpinski18}. However, a complete prescription for manipulating temporal modes using linear-optical methods has yet to be developed. \par

Recent experiments with phase-only manipulation of optical pulses have demonstrated spectral shearing \cite{fan16, wright17} and bandwidth manipulation \cite{karpinski18}. These techniques have been developed using the optical space-time analogy \cite{TORRES}, in which the evolution of a pulsed and transverse-spatial modes can be mapped onto equivalent equations of motion. Examples include the time lens \cite{kolner89,kolner94} and the dispersive Fourier transform \cite{avenhaus09,TORRES,Goda}. Both of these applications have been implemented using off-the-shelf components. \par

In this paper we build on the previous work and present a general recipe for implementing targeted unitary transformations on temporal modes through the use of temporal phase modulation and spectral dispersion.
First we introduce the mathematical framework by which these transformations can be described. Then we introduce an experimental apparatus that can perform these transformations based upon electro-optic phase modulation and chirped fiber Bragg gratings. Numerical simulations are then presented that show the performance of the proposed experimental apparatus. The simulations, performed on select single-mode and multi-mode transformations applicable to quantum applications, take into account realistic specifications for the modulators and gratings.\par

\section{Theory}
To describe the quantized electromagnetic field one must first decompose the field into an appropriate set of orthonormal field modes that satisfy the Maxwell equations \cite{mandel-wolf}. Here we choose to focus on a paraxial geometry, such as found within integrated optical systems or in beam-like propagation directed along the $z$-axis. In addition, we assume pulsed modes traveling with group velocity $v_g$ in which the slowly-varying envelope approximation holds, that is, where the temporal envelope of the pulse changes on a time scale significantly longer than an optical-carrier cycle period, $T_0=2\pi/\omega_0$. In these limits the electric field vector can be expressed as 
\begin{equation}
{\bf{E}}({\bf{r}},t) = \sum_{m}\alpha_{m} {\bf{e}}_{m} u_{m}({\bf{r}}) \psi_{m}(t-z/v_g) e^{i(k_{0}z-\omega_{0} t)}.
\end{equation}
Here the summation is taken over a set of discrete, orthonormal modes describing the transverse spatial, $\{u_{m}({\bf{r}})\}$, temporal, $\{\psi_{m}(t-z/v_g)\}$, and polarization, $\{{\bf{e}}_{m}\}$, degrees of freedom of the field. These are weighted with a complex amplitude, $\alpha_{m}$, and have associated wave vector, $k_{0}$, and carrier frequency, $\omega_{0}$, which are all labelled by a non-negative integer $m$. Here we focus on the temporal modes $\{\psi_{m}(t-z/v_g)\}$ of the field and consider the case in which only a single polarization $\bf{e}$ and transverse spatial mode $u({\bf{r}})$ are excited. Dropping the polarization and transverse spatial mode dependence, the electric field in the time domain is thus
\begin{equation}
E(z,t) = \sum_{m} \alpha_{m} \psi_{m}(t-z/v_g) e^{i(k_{0}z-\omega_{0} t)}.
\end{equation}

In the following, we will work in the reference frame of the pulse traveling at the group velocity, $v_g$, along the $z$-axis, and thus drop the $z$-dependence of the temporal modes. The temporal modes form a complete set that can be used to expand an arbitrary pulsed signal and obey the orthonormality relation
\begin{equation}
\int_{-\infty}^{+\infty} \psi_{m}^{*}(t)\psi_{n}(t){\rm{d}}t = \delta_{mn}.
\end{equation}
Note that a temporal mode $\psi_{m}(t)$ can also be represented by its Fourier transform, ${\tilde{\psi}}_{m}(\omega)$, in the frequency domain, where the orthonormality condition still holds
\begin{equation}
\int_{-\infty}^{+\infty} {\tilde{\psi}}_{m}^{*}(\omega){\tilde{\psi}}_{n}(\omega){\rm{d}}\omega = \delta_{mn}.
\end{equation}

Standard techniques to control the pulse mode of light involve either filtration or amplification in the frequency or time domain \cite{diels-rudolph, weiner11}. Such an approach to pulse shaping is incompatible with quantum states of light owing to added noise and reduction in the signal arising from amplification and loss. These lead to the destruction of the fragile quantum coherences between different photon-number components of the state. To preserve the quantum state of light occupying different temporal modes when modifying the mode structure thus requires unitary, i.e. phase-only, manipulation. In practice, unitary mode transformations are challenging to achieve due to technical losses. Still, deterministic control of optical pulsed modes, in which each pulse successfully passing through a device is precisely modified, can be attained by application of spectral and temporal phase, which is then only reduced in amplitude by an overall efficiency factor.

A simple example of how application of phase can modify the amplitude distribution of a pulse is the time domain analogy of Fraunhofer diffraction, known as the dispersive Fourier transform \cite{TORRES,Goda}. Consider a transform-limited pulse $\psi(t)$ and its Fourier transform $\tilde{\psi}(\omega)$. After propagating through a medium with second-order dispersion characterized by the group delay dispersion (GDD) $\delta$, which applies a quadratic spectral phase to the pulse, the spectral mode amplitude is
\begin{equation}
\tilde{\psi}'(\omega) = \tilde{\psi}(\omega)e^{i \delta (\omega-\omega_{0})^{2}/2},
\label{2ODF}
\end{equation}
where $\omega_{0}$ is the central frequency of the pulse \cite{diels-rudolph, weiner}. The amplitude of the pulse in the time domain becomes a convolution of the original temporal mode with the Fourier transform of the quadratic spectral-phase factor and can be expressed as
\begin{equation} 
\psi'(t) = Ne^{-it^2/2\delta}\int_{\infty}^{-\infty}\psi(\tau)e^{-i\tau^2/2\delta}e^{2it\tau/\delta} {\rm{d}}\tau, 
\label{2ODT}
\end{equation}
where $N$ is a normalization constant. If the temporal duration of the pulse $\psi(t)$, given by $\tau_p$, is short compared with the square root of the GDD, i.e. $\tau_p << \sqrt{\delta}$, then the Gaussian temporal phase factor in Eq. (\ref{2ODT}) can be treated as unity within the integral and the resultant temporal mode at the output is just a Fourier transform of the input,
\begin{equation}
\psi'(t) = Ne^{-it^2/2\delta}\tilde{\psi}\left(\frac{2t}{\delta}\right).
\label{DFT}
\end{equation}
The frequency argument of the Fourier transform is $\omega = {2t}/{\delta}$. By the analogy to spatial diffraction this is called the temporal far-field condition and is satisfied approximately when
\begin{equation}
\frac{\tau_p^2}{\delta} < \frac{\pi}{16}.
\label{eq:farfield}
\end{equation}

Thus by introducing sufficient group-delay dispersion to satisfy the temporal far-field condition of Eq. (\ref{eq:farfield}) the temporal profile of the pulse becomes a scaled replica of the spectrum. The dispersive Fourier transform is a phase-only operation and thus a unitary transformation of the field modes, which makes it compatible for use with quantum states of light. Indeed, this technique has been used to measure the spectrum of single-photon pulses using highly-dispersive optical fibers \cite{avenhaus09} and chirped-fiber Bragg gratings \cite{davis17} along with time-resolved single-photon detection.\par

Inspired by methods to implement programmable unitary transformations on transverse spatial modes, demonstrated through the use of sequential application of spatial phase and diffraction \cite{Morizur}, we present here an approach to realize temporal mode unitary transformations by application of temporal phase and dispersion. This approach follows from the fact that unitary transformations can be decomposed into a sequence of applied phases and Fourier transforms \cite{Borevich, SCHMID}. In practice, we consider discretized field modes $\psi_i=\psi(t_i)$ that are sampled versions of the mode functions at discrete times $\{t_i\}$, with $i$ a positive integer labeling sample time. In this case the continuous modes become vectors and the unitary transformations become unitary matrices. The unitary matrices can then be decomposed into a sequence of diagonal unitary matrices and discrete Fourier transforms \cite{SCHMID}. For a given unitary matrix $U$ its decomposition can be written as
\begin{equation} \label{eq:unitarydecomposition}
U = \mathfrak{F}^{\dagger}D_{N}\cdots \mathfrak{F}D_{2}\mathfrak{F}^{\dagger}D_{1},
\end{equation}
where the $D_{j}$ are diagonal unitary matrices, $\mathfrak{F}$ is the discrete Fourier transform, and $\mathfrak{F}^{\dagger}$ is its inverse. This decomposition is at the heart of our approach to implementing programmable unitary temporal mode transformations. It requires the ability to perform sequential application of phases in the temporal and spectral domains. Here we propose a method to implement such temporal mode unitary transformations using a sequence of dispersive Fourier transforms and application of temporal phase profiles.\par

\section{Proposed Implementation}
To implement a targeted temporal mode transformation $U$ a sequence of phases can be achieved by application of temporal phase followed by the dispersive Fourier transform, which maps the spectral amplitude of a pulse onto its temporal envelope as in Eq. (\ref{eq:farfield}). As the spectrum has been mapped onto the temporal envelope, application of a time-varying phase implements a spectral phase to the original pulse. This is followed by the inverse dispersive Fourier transform, which can be achieved using a medium with the same magnitude GDD, but opposite sign.\par

Here we consider temporal phase modulation implemented using an electro-optic phase modulator (EOPM). The phase applied to an optical field traveling through an EOPM, $\phi(t)$, is determined by the electronic voltage, $V(t)$, that drives the modulator and the intrinsic electro-optic response of the EOPM itself. The applied phase is proportional to the driving voltage, $\phi(t)=V(t)/V_{\pi}$, given that the driving signal does not exceed the electronic bandwidth of the modulator, which we denote $B$. Here $V_{\pi}$ is the voltage required to achieve a $\pi$ phase shift at the highest operating frequency of the modulator. Therefore the temporal phase profiles that can be apply are restricted by the capabilities of the EOPM, encapsulated by $V_{\pi}$ and $B$, and the electronics available that generate and deliver the driving voltage $V(t)$. 

The phase modulation bandwidth $B$ restricts the frequencies of the phase profiles that can be applied. Traveling-wave EOPMs are capable of modulation bandwidths in the tens of gigahertz region with low $V_{\pi}\approx 2$ V. State-of-the-art modulators have demonstrated electronic bandwidth up to $B \approx 100$ GHz with $V_{\pi}=2.3$ V \cite{Wang:18}. Similarly, arbitrary waveform generators for radio-frequency (RF) generation can produce electronic signals with up to $100$ GHz bandwidth and $2$ V amplitude. Thus in the following we consider temporal mode transformations composed from temporal phase modulation with phase modulation bandwidths in the $1-100$ GHz region.

The dispersive Fourier transform in Eq. (\ref{eq:farfield}), which maps the spectral amplitude onto the temporal amplitude of the pulse, is performed by applying GDD, $\delta$, a quadratic spectral phase, to the pulse. This can be applied by propagating through a length of optical fiber. The length required is proportional to the square of the temporal duration of the pulse. Performing a Fourier transform on a pulse with a temporal duration of $\tau_p=10$ ps with $800$ nm central wavelength would require $10$ km of optical fiber. Depending on the wavelength regime, a fiber of this length would introduce significant loss, which cannot be tolerated for unitary transformations. Another approach to achieve second-order spectral dispersion is by using a chirped fiber Bragg grating (CFBG), which can apply a large amount of dispersion with relatively low loss compared with an optical fiber. CFBGs can be manufactured with quadratic dispersion equivalent to that achieved in over $10$ km of fiber. In addition, a CFBG can easily implement both positive and negative GDD by simply reversing the direction into which light is launched into the CFBG. In either case, however, there is a finite limit to the GDD, $\delta$, that can be implemented using a fiber or CFBG, which effectively places an upper limit on the temporal duration of the pulses used. \par

The bandwidth constraint of the EOPM and the upper limit on GDD determines the range of temporal widths of the pulses that should be used. For instance consider an EOPM with $B=50$ GHz bandwidth. According to the sampling theorem we can generate a temporal phase profile by defining samples at approximately $10$ ps intervals. This places a lower limit on the temporal width of optical pulses that can be addressed. In general, a transformation on $N$ temporal modes requires pulses to have sufficient duration so that $N$ samples of the phase modulation can be applied across them so that there are enough independent degrees of freedom to operate on each mode.\par
The number of samples applied to the spectral phase must also be considered. For a Gaussian pulse with a temporal duration $\tau$ the spectral width is $\nu = 8\ln(2)/\tau$. So after dispersive Fourier transform, by Eq. (\ref{DFT}), the pulse will have a temporal duration $\tau_{\omega} = 4\ln(2)\delta/\tau$. The total number of phase samples we achieve in the spectral domain with this approach is proportional to $\tau_{\omega} + \tau$, which increases as $\tau$ gets smaller. This implies that our approach is limited to pulses whose transform-limited durations cover at least $N$ temporal phase samples of the phase modulation to ensure the ability for universal pulse mode transformation.

\subsection{Proposed Experiment}
\begin{figure}[h] 
\centering
  \includegraphics[scale=0.87]{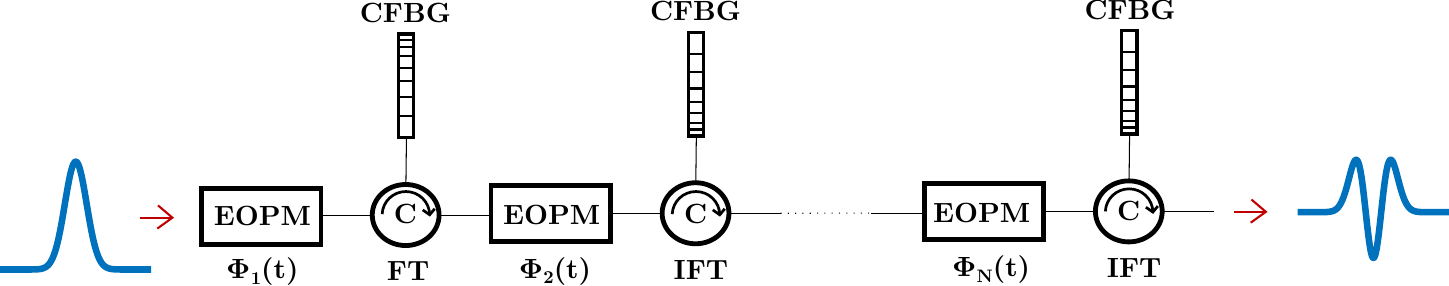}
  \caption{Proposed experimental setup. A sequence of $N$ sets of electro-optic phase modulators (EOPM), circulators (C) and chirped-fiber Bragg gratings (CFBG), the latter with alternating signs of GDD, implement a series of $N$ phase profiles $\Phi_{j}(t)$, $j=1,\cdots, N$, where $N$ is the total number of steps in the decomposition of Eq. (\ref{eq:unitarydecomposition}). Each phase profile implements a diagonal element of the decomposition, $D_j$ in Eq. (\ref{eq:unitarydecomposition}). Chirped-fiber Bragg gratings (CFBG) are used to perform the Fourier transform (FT) and inverse Fourier transform (IFT).}
\label{fig:setup}
\end{figure}

A unitary temporal mode transformation decomposed as in Eq. (\ref{eq:unitarydecomposition}) can be implemented using the experimental setup depicted in Fig. \ref{fig:setup}. The apparatus consists of a series of EOPMs, circulators (C) and CFBGs. A pulse enters through the first phase modulator, which is driven by a shaped electronic signal such that the targeted phase, $\Phi_1(t)$, that realizes the diagonal element of the decomposition, $D_1$, in Eq. (\ref{eq:unitarydecomposition}) is applied. The pulse then passes through a circulator and reflects off a CFBG with positive GDD, $\delta$, performing the Fourier transform (FT). The output of the circulator leads to another phase modulator which is driven to apply the second targeted phase, $\Phi_2(t)$, that realizes the diagonal element of the decomposition, $D_2$. The pulse continues through a second circulator and CFBG, the latter of which is oriented to apply negative GDD, $-\delta$, performing the inverse Fourier transform (IFT). This process is repeated, with the pulse passing through as many sets of phase modulators and CFBGs as needed according to the decomposition. Alternatively, one could achieve the same transformation by switching the pulse into a cavity comprised of two CFBGs with opposite signs as the mirrors and an EOPM inside the cavity.

\section{Simulation}

Although Eq. (\ref{eq:unitarydecomposition}) indicates that any unitary transformation on the temporal modes has a decomposition as a sequence of temporal phase modulations and Fourier transforms, it does not provide a way to determine the form of the phases required or the number of factors in the decomposition of a given transformation. Furthermore, we are not able to implement arbitrary temporal phase profiles since the bandwidth and driving voltage of the EOPMs are constrained. To demonstrate that such a unitary decomposition works in a practical setting, we present numerical simulations that determine an optimal set of phases for a fixed number of phase factors within the sequence to implement select targeted unitary transformations. These simulations takes into account the finite bandwidth of the EOPMs as well as the driving electronics. 

A unitary transformation $U$ can be represented in various bases. Here we will work in either the time basis or the basis of Hermite-Gaussian (HG) modes defined by

\begin{equation}
HG_{n}(t) = \frac{1}{\sqrt{2^{n}n!\sqrt{2\pi}\sigma}}H_{n}\left(\frac{t}{\sqrt{2}\sigma}\right)e^{-\left(\frac{t}{2\sigma}\right)^2},
\end{equation}
where $H_{n}(t)$ is the $n$-th order Hermite polynomial. The full-width at half-maximum (FWHM) duration of the fundamental Gaussian pulse is $\tau_p=2.35\sigma$. This set of modes forms a complete orthonormal basis for the temporal modes. 

Each simulation begins with a targeted transformation, $U$, which we express in the HG basis. For example $U_{i,j}$ represents the component of the input temporal mode that overlaps with the $j^{\rm{th}}$ HG mode, $HG_j$, that is mapped to the $i^{\rm{th}}$ output HG mode, $HG_i$. A unitary transformation is completely determined by its action on a complete set of input modes. Thus, for a given simulation we calculate the output of the realistic transformation, $T$, when acting on the set of input HG modes $\{HG_{m}(t)\}$, which gives output modes $\{\chi_m(t) = T \cdot HG_{m}(t)\}$. The output modes are then compared with the targeted output modes $\{\xi_{m}(t) = U\cdot HG_{m}(t)\}$, with $m=1,2,\cdots$, that would be the ideal output of the transformation. 

To find an optimal realistic transformation $T$ for a fixed number of phase applications, denoted $N$, constrained by finite electronic bandwidth, $B$, and half-wave voltage, $V_{\pi}$, of the EOPMs and driving electronics and the limited dispersion available, $\delta$, we use a modified simulated-annealing algorithm. Each simulation begins with discretization of the input $\{HG_{m}(t)\}$ and target $\{\xi_{m}(t)\}$ modes, which have transform limited duration of approximately $\tau_{p} \approx M/2B$, where $M$ is the number of modes involved in the target transformation. This choice of $\tau_{p}$ is to ensure that the modes are wide enough to be independently modulated. For each simulation the modes are defined on a temporal window of $1$ ns, chosen to be several times wider than the longest pulses so that any temporal shifts or temporal broadening caused by the spectral phase modulation will remain within this window. They are sampled at a rate higher than twice the spectral width of the modes to prevent aliasing. The phase modulations, $\Phi_n(t)$ are initially generated by random samples in the range $[0,\pi]$ at the frequency $2B$ corresponding to the phase modulation bandwidth $B$. This ensures that the generated phases can reach $V_{\pi}$ while remaining within the electronic bandwidth constraints. They are then interpolated to the same sampling rate as the modes. The dispersive Fourier transform is simulated by performing a fast Fourier transform (FFT) on the modes, multiplying each mode by a quadratic spectral phase, and then performing an inverse FFT. \par

The set of input modes is operated on alternately by the phase modulations and dispersive Fourier transforms up to a predefined number of steps, $N$. After the final step the transformed input modes, $\{\chi_{m}(t) = T \cdot HG_{m}(t)\}$, are compared to the target modes, $\{\xi_{m}(t) = U \cdot HG_{m}(t)\}$, using a cost function tailored to the type of transformation performed, either a one-to-one mode mapping or $M$-to-$M$ mode mapping as described below. This cost function is minimized using a simulated annealing algorithm to optimize the phases.\par

\subsection{Single-mode Transformation}
When transforming one input pulse, $\psi(t)$, to an output targeted pulse, $\xi(t)$, there are, in principle, an infinite number of unitary transformations that can achieve this objective. Our simulations on such single-mode transformations aim to determine the $N$ applied phases $\{\Phi_j(t)\}$, $j=1,\dots,N$, subject to experimental constraints that optimizes the overlap of the output and target modes. The simulated-annealing algorithm initializes the set of $N$ phases, $\{\Phi_j(t)\}$, by randomly setting each phase sample in the range $[0,\pi]$. From these phases the transformation $T=\mathfrak{F}^{\dagger}D_{N}\cdots \mathfrak{F}D_{2}\mathfrak{F}^{\dagger}D_{1}$ is computed. This transformation is then applied to the input mode, $\psi(t)$ to give an output mode, $\chi(t)=T\cdot\psi(t)$. We use the fidelity as a measure of closeness between the output and target modes, defined as the modulus squared of the overlap of the output mode with the target mode
\begin{equation}
F =  \left|\int \chi^{*}(t)\xi(t) \mathrm{d}t  \right|^2.
\label{eq:fidelity}
\end{equation}
The algorithm generates an additional set of phases $\{\bar{\Phi}_j(t)\}$, $j=1,\dots,N$ by randomly perturbing a small number of samples of from the phases in the original set. The corresponding transformation $\bar{T}=\mathfrak{F}^{\dagger}\bar{D}_{N}\cdots \mathfrak{F}\bar{D}_{2}\mathfrak{F}^{\dagger}\bar{D}_{1}$ is calculated along with the output mode $\bar{\chi}(t)=\bar{T}\cdot\psi(t)$ and fidelity $\bar{F}$. The algorithm starts with a temperature $K$. In each iteration the new set of phases is compared to the prior set of phases by computing $\Delta F = \bar{F}-F$. The new set of phases is taken with a probability $P(\Delta F)=e^{\frac{\Delta F}{K}}$, otherwise the initial phase settings are kept. The temperature is lowered with each iteration.
The simulation results in a fidelity $F \leq 1$, where the maximum $F = 1$ would be the fidelity of the ideal transformation. For single-mode transformations we use this fidelity as the cost function, so the simulation maximizes the fidelity. Note that $L=1-F$ gives the loss associated with the transformation into modes orthogonal to the target mode. \par

\begin{figure}[H] 
\centering
  \includegraphics[scale = 0.9]{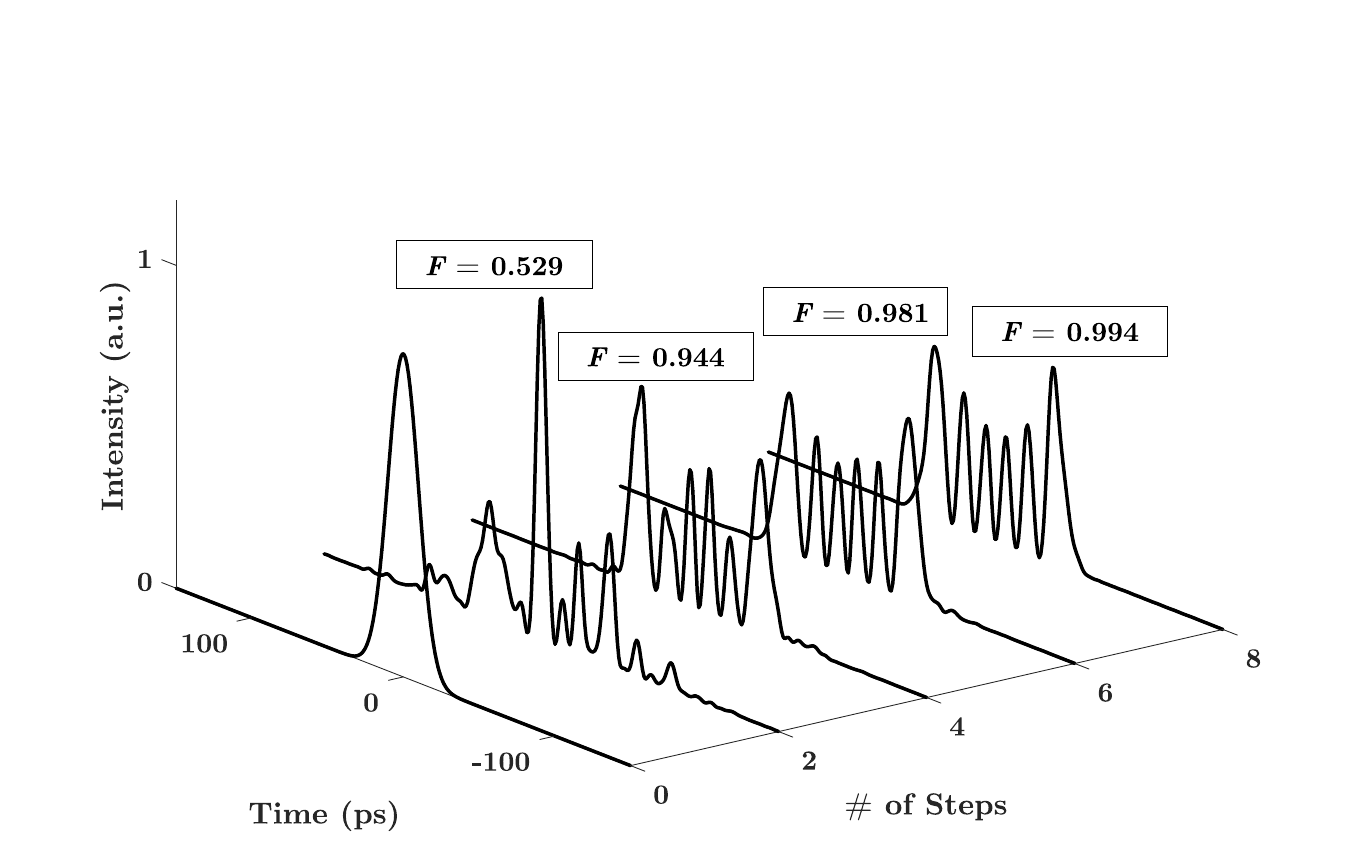}
  \caption{Temporal intensity profile, $I(t)=|\chi(t)|^2$, resulting from the simulated transformation of a Gaussian temporal mode into the $5^{\rm{th}}$-order HG mode each having Gaussian temporal duration parameter with $\sigma = 10$ ps. The simulations assume a phase modulation bandwidth of $B=40$ GHz, optimized over a range from $0$ steps (left) to $8$ steps (right).}
\label{fig:hg5sim}
\end{figure}

As an example we simulate the transformation of a $0^{\rm{th}}$-order HG mode with $\sigma = 10$ ps to a $5^{\rm{th}}$-order HG mode with the same $\sigma$ value. The phase modulation bandwidth is restricted to $B=40$ GHz, and the simulation is performed for several different step numbers, $N$. The results shown in Fig. \ref{fig:hg5sim} depict the output pulse intensity profile, $I(t)=|\chi(t)|^2$ for phase steps of $N=2,4,6,8$. We see that the fidelity increases with step number, reaching $F=0.994$ after $8$ steps. Note that only after four phase settings, which implies two sets of combined spectral and temporal phase applications, nearly $95$\% fidelity can be achieved.

\begin{figure}[H]
\centering
  \includegraphics[scale=0.85]{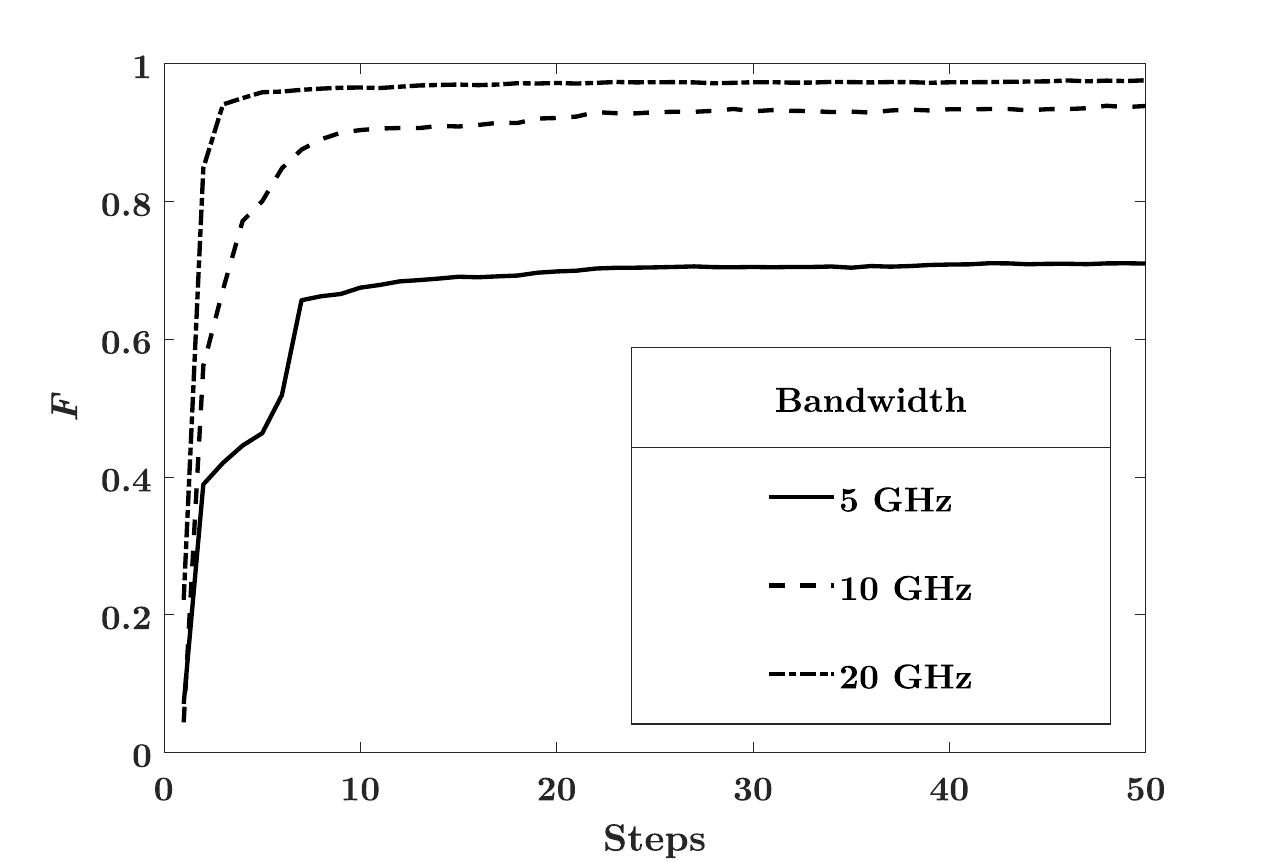}
  \caption{The fidelity of the optimized $HG_{0}$ to $HG_{5}$ transformation, $F$, as a function of the number of phase modulation steps, $N$. The fidelity is plotted for phase modulation bandwidths of $5$ GHz, $10$ GHz, and $20$ GHz and fixed pulse duration parameter, $\sigma = 10$ ps. Each transformation converges to a nearly constant fidelity, within $8$ steps for the different modulation bandwidths. The higher bandwidth modulator converges in fewer steps and to a larger fidelity than the lower bandwidth modulators.}
   \label{fig:converg}
\end{figure}

 To determine how the fidelity behaves with the number of steps we simulate the same $HG_0$ to $HG_5$ mode transformation over large range of steps for modulation bandwidths of $B=5, 10, 20$ GHz, keeping the pulse duration parameter, $\sigma = 10$ ps, fixed. The results, shown in Fig. \ref{fig:converg}, indicate that for a fixed modulation bandwidth and pulse duration fidelity plateaus and additional phase modulation steps do not significantly lead to increased fidelity. We also see that increased bandwidth leads to higher achievable fidelity and convergence to this fidelity in fewer steps. This leads one to conjecture that there are some transformations that are infeasible under realistic constraints. Furthermore, this indicates that there is a trade off in the resources needed for an experiment, quantified by the modulation bandwidth available and the number of phase steps required to achieve a target transformation.

\subsection{Multi-mode transformations}

Now we turn our attention toward multi-mode transformations. Here we consider $M \times M$-dimensional transformations acting on a basis of $M$ HG modes. The transformation $U$ maps $M$ input HG modes onto a set of $M$ orthonormal target output modes $\{\xi_m(t) = U \cdot HG_m(t)\}$. We simulate this transformation using a similar procedure as used in the single-mode case. For a fixed number of phase modulation steps, $N$, the set of $N$ phases, $\{\Phi_j(t)\}$, is again initialized by randomly setting each phase in the range $[0,\pi]$. The realistic transformation $T=F^{\dagger}D_{N}\cdots FD_{2}F^{\dagger}D_{1}$ is calculated using the dispersive Fourier transform and is applied to each of the $M$ input modes simultaneously. We can define an $M$-by-$M$ matrix $U$ for the target transformation on the $M$ HG modes, where the matrix elements in the HG-mode basis are given by the overlap of the target modes with the HG modes

\begin{equation} \label{eq:mmtelem}
U_{ij} = \int HG_{i}^{*}(t)\xi_{j}(t)\mathrm{d}t.
\end{equation}
%Note that this assumes the input modes are Hermite-Gaussian modes, $\{\xi_j(t)\}=\{HG_j(t)\}$. 
Similarly, we define a matrix $T$ for the output of the simulation where the output modes of the transformation, $\{\chi_j(t)\}=\{T \cdot HG_j(t)\}$ replace the target modes in Eq. (\ref{eq:mmtelem}), giving
\begin{equation} \label{eq:mmtelem2}
T_{ij} = \int HG_{i}^{*}(t)\chi_{j}(t)\mathrm{d}t.
\end{equation}

To evaluate the effectiveness of the transformation generated by this procedure we define the fidelity as
\begin{equation}
    F = \frac{1}{M^2}\left|{\rm{Tr}}(T^{\dag}U)\right|^2.
\end{equation}
where ${\rm{Tr}}(\cdot)$ is the trace operation. Note that while the trace is only over the $M$ HG modes the transformation $T$ may send energy into modes outside this $M$-mode subspace. To quantify this we introduce two new quantities related to the fidelity by
\begin{equation} \label{eq:f}
    F = P\times F_{HS}.
\end{equation}
Where $P$ is the success probability
\begin{equation} \label{eq:P}
P = \frac{1}{M}{\rm{Tr}}\left(T^{\dag}T\right).
\end{equation}
This quantifies the probability that the output modes resulting from the transformation remain in the chosen $M$-mode subspace. A transformation that has a value of $P < 1$ means that there is loss into the other modes. The other quantity used to evaluate the simulated transformation is the Hilbert-Schmidt fidelity \cite{WANG}
\begin{equation} \label{eq:fHS}
F_{HS} = \frac{1}{M^2}\frac{\left|{\rm{Tr}}(T^{\dag}U)\right|^2}{P}.
\end{equation}
This can be interpreted as the probability that, given the output modes stay in the target subspace, the desired transformation is implemented. An optimal transformation would result in $P = 1$ and $F_{HS} = 1$. \par
After applying the phase modulations to the input pulses we calculate the matrix elements of the transformation as in Eq. (\ref{eq:mmtelem2}). The phases are then optimized to maximize the fidelity, Eq. (\ref{eq:f}), using the iterative simulated annealing algorithm. We then compute the success probability and Hilbert-Schmidt fidelity. Here we consider three targeted multi-mode transformations.\par
\subsubsection{Hadamard transformation}
\begin{figure}
\centering
\begin{subfigure}{.5\textwidth}
  \centering
  \includegraphics[width=\linewidth]{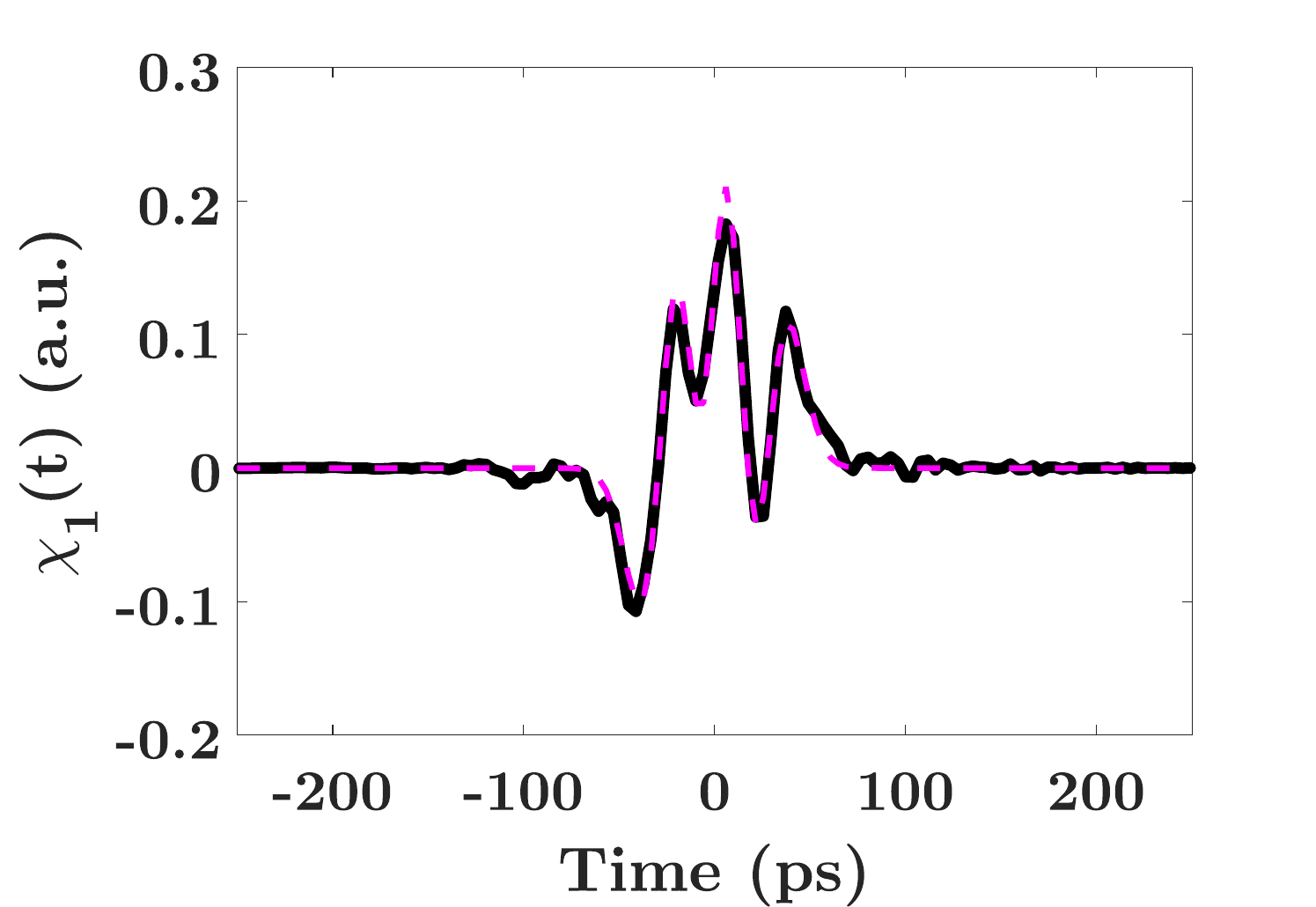}
  \caption{Output mode $\chi_{1}(t)$}
  \label{fig:sub1}
\end{subfigure}%
\begin{subfigure}{.5\textwidth}
  \centering
  \includegraphics[width=\linewidth]{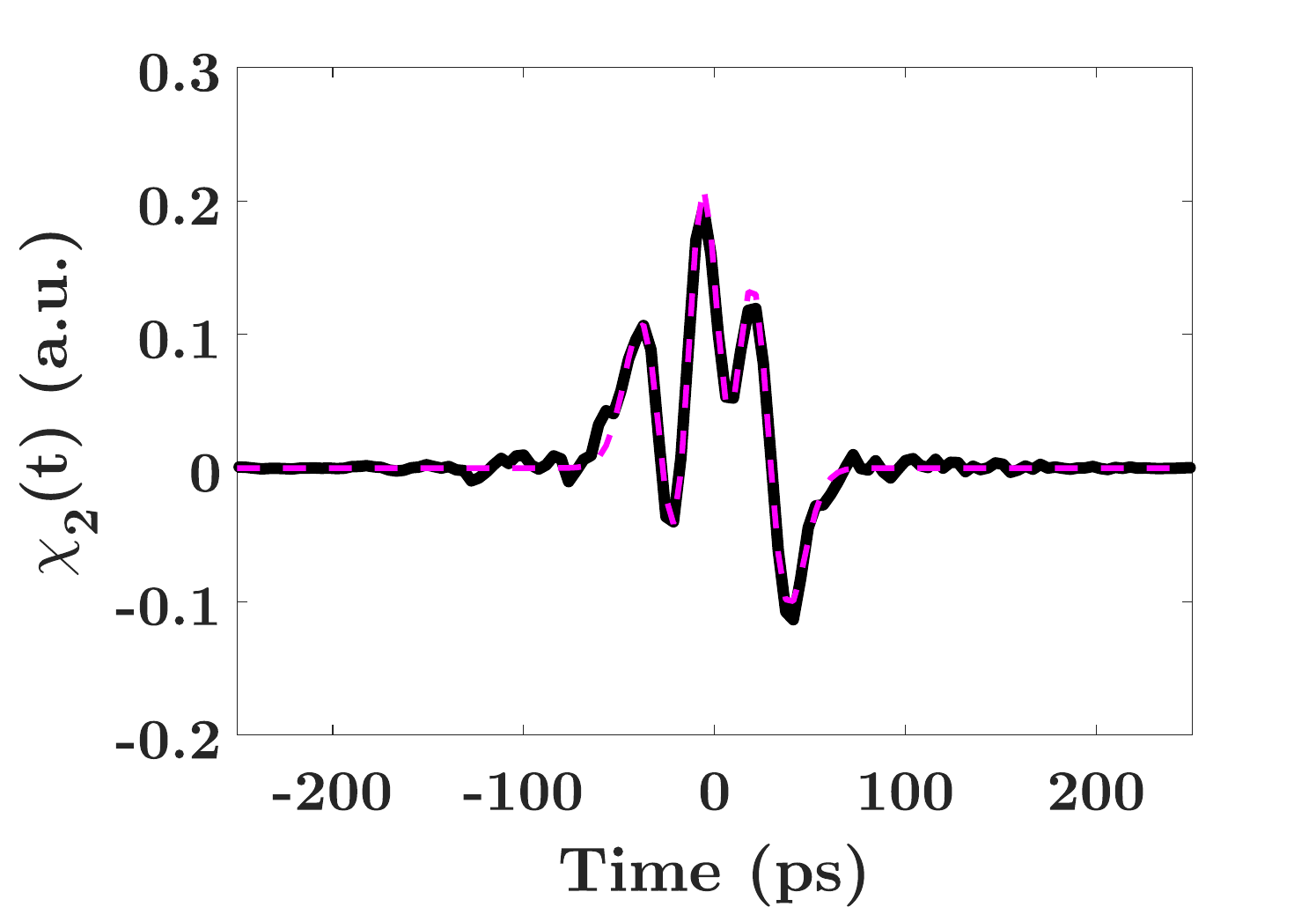}
  \caption{Output mode $\chi_{2}(t)$}
  \label{fig:sub2}
\end{subfigure}
\caption{The temporal amplitude of the output modes of the Hadamard simulation. The simulation assumes a phase modulation bandwidth of $B = 20$ GHz, optimized over $N=6$ phase steps. The temporal amplitude of the target modes are displayed as dashed lines.}
\label{fig:hadplot}
\end{figure}

To begin with we demonstrate the ability of this system to implement a beam splitter, or Hadamard, transformation between two HG modes, which is represented by a matrix in the space of the two HG modes as
\begin{equation}
U = \frac{1}{\sqrt{2}}\begin{pmatrix}
1 & 1 \\
1 & -1
\end{pmatrix}.
\label{eq:hadamard}
\end{equation}
For this simulation we use the $0^{\rm{th}}$- and $5^{\rm{th}}$-order HG modes as our two-mode subspace, with equal pulse duration parameter, $\sigma = 10$ ps. The applied phases, subject to a modulation bandwidth constrained to $B=20$ GHz, are optimized to maximize the Hilbert-Schmidt fidelity $F_{HS}$. The temporal mode outputs for each input resulting from this simulation using $N=6$ steps are depicted in Fig. (\ref{fig:hadplot}). \par

The output modes of the simulation closely match the target modes of the desired transformation with overlap fidelities $F_{1} = 0.923$ and $F_{2} = 0.952$. The presence of rapid oscillations outside the main peaks of the amplitudes in Fig. (\ref{fig:hadplot}) can be considered as loss into higher-order HGn modes outside of the two-mode subspace. This contributes to a success probability $P=0.941$ that is less than the ideal value of $1$. The output matrix $T$ of this transformation is 
\begin{equation}
T = \frac{1}{\sqrt{2}}\begin{pmatrix}
0.96 - 0.10i & 0.98 - 0.03i \\
0.96 - 0.02i & -0.97 - 0.01i
\end{pmatrix}
\end{equation}
The Hilbert-Schmidt fidelity of this matrix with the target in Eq. (\ref{eq:hadamard}) is $F_{HS} = 0.999$. The success probability $P=0.941$ indicates loss into the other HG modes, but the high Hilbert-Schmidt fidelity indicates that within the two-mode subspace the transformation performs well.
\subsubsection{Four mode Hadamard transformation}
In principle, we should be able to perform targeted transformations on an arbitrarily large number of modes. However, we have seen that experimental constraints restrict the types of transformations that can be performed and may also impact the number of modes that can be addressed in a given experimental configuration. As an example we simulate a four-mode Hadamard transformation between the first four HG modes defined by the matrix
\begin{equation}
U =  \frac{1}{2}\begin{pmatrix}
1 & 1 & 1 & 1 \\
1 & -1 & 1 & -1 \\
1 & 1 & -1 & -1 \\
1 & -1 & -1 & 1
\end{pmatrix} .
\label{eq:hadamard4}
\end{equation}
This transformation is simulated using a $B=20$ GHz bandwidth modulator and $N=6$ phase steps. The output matrix $T$ from the simulation was
\begin{equation}
T =  \frac{1}{2}\begin{pmatrix}
0.67 + 0.07i & 1.00 + 0.04i & 1.10 + 0.01i & 0.80 - 0.47i \\
1.01 + 0.00i & -0.98 - 0.02i & 0.75 - 0.07i & -0.85 + 0.12i \\
0.98 - 0.03i & 0.98 + 0.08i & -0.91 + 0.07i & -0.89 + 0.07i \\
0.95 + 0.01i & -0.68 - 0.05i & -0.86 + 0.01i & 1.16 - 0.11i
\end{pmatrix} 
\end{equation}
From this matrix we compute a Hilbert-Schmidt fidelity with the target matrix in Eq. (\ref{eq:hadamard4}) of $F_{HS} = 0.975$ and a success probability $P = 0.852$. Further simulations for higher-dimensional generalized Hadamard transformations showed that as the number of modes increases, so do the number of steps required to reach a maximum fidelity, which was lower under the same constraints. \par

\subsection{Demultiplexer}
\begin{figure}[H]
\centering
  \includegraphics{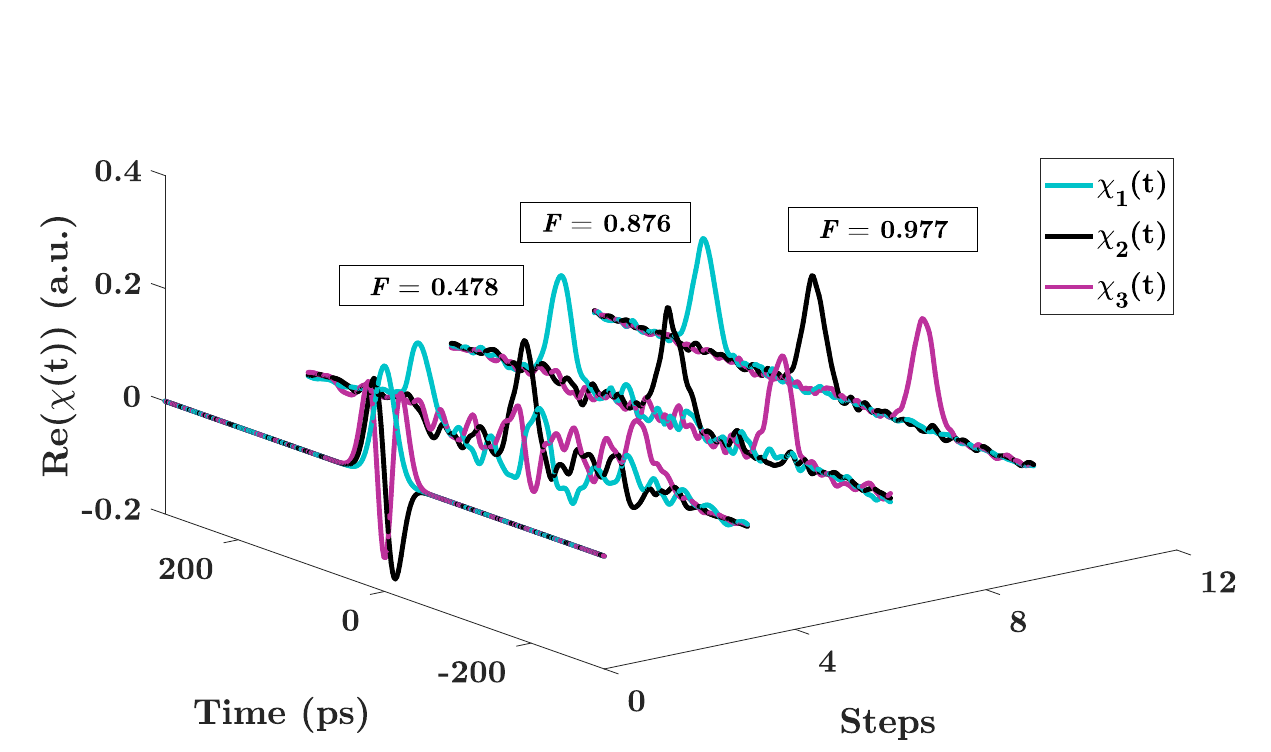}
  \caption{Demultiplexing the temporally overlapped Gaussian (turquoise), a $1^{\rm{st}}$-order HG (black), and $2^{\rm{nd}}$-order HG (magenta) pulses into temporally separated Gaussian pulses for an increasing number of phase modulation steps. The initial temporally overlapping HG pulses and target pulses have widths $\sigma = 20$ ps. The target Gaussian modes are separated by $200$ ps. The average fidelity is shown for each transformation. }
\label{fig:demux}
\end{figure}
As a final application of this system we look at a temporal-mode demultiplexer, which sorts HG modes centered at one time into Gaussian pulses temporally separated such that a fast detector could distinguish them by time of arrival.
By separating the HG-modes into separate time bins the information encoded in the temporally-overlapping HG modes can be read out using a time-resolved detector. We simulate this operation on the first three HG modes with a temporal duration parameter $\sigma = 20$ ps, demultiplexing them into individual Gaussian modes separated in time by $200$ ps. For this simulation we use the fidelity of Eq. (\ref{eq:fidelity}) and optimize the set of applied phases to maximize the average fidelity of the three modes. The results of this simulation, shown in Fig. \ref{fig:demux}, for $N=0,4,8,12$ phase steps. To evaluate the efficacy of this transformation for demultiplexing we calculate the average signal-to-noise ratio (SNR) across the time bins. The SNR for each time bin given by the ratio of the energy in each time bin arising from the targeted mode to the energy in each time bin from the other two modes. For this transformation we find an SNR of $187$, $66$, and $88$ for modes $0$, $1$, and $2$ respectively. We see that this system is capable of performing the demultiplexing operation on three modes. After 12 steps the modes are demultiplexed with an average fidelity of $F = 0.978$.

\section{Conclusion}
 A method to implement temporal mode transformations that are, in principle, unitary operations has been introduced. This has been shown to be experimentally feasible with current technology by means of numerical simulations. The approach uses a decomposition of the mode transformations into into a sequence of temporal phase modulations and Fourier transforms. The fidelity that can be achieved in our simulations is limited by the experimental constraints, including the electronic bandwidth of the EOPMs and available amount of optical dispersion. Under these constraints we have shown that transformations on up to four temporal modes with can be performed with fidelity greater than $95\%$. The number of steps needed to perform these transformations is proportional to the number of modes. \par

The simulations here neglected loss, which suffices for many circumstances. In principle, since each step to generate a temporal mode transformation involves phase-only operations the overall transformation should be lossless. However, in a realistic experimental setup each component would contribute to loss in the system. Furthermore, frequency-dependent loss will play a significant role in the design of any device acting on temporal modes. \par
A further experimental challenge to be overcome is the generation of the electronic signals needed to drive the EOPMs. Each phase modulator must be driven by a different specified electronic signal for each step in the decomposition, and these must be synchronized with the arrival of the pulse at the EOPMs. Novel schemes to solve this problem have recently been demonstrated by using photonic waveform generation \cite{karpinski20}. \par
The implementation of this system would enable targeted multimode unitary transformations on temporal modes. This would allow the use of the temporal modes as a basis for encoding quantum information with wide ranging applications from computing and communications to dynamic spectroscopy and sensing. \par

\section*{Funding}
National Science Foundation (1620822); Defence Science and Technology Laboratory (DSTLX-100092545); Horizon 2020 Framework Programme (665148).

\section*{Acknowledgments}
The authors thank Micha\l{} Karpi\'{n}ski, Michael Raymer and Nicolas Treps for helpful discussions.

\section*{Disclosures}
The authors declare no conflicts of interest.

\bibliography{bibliography} 

\begin{thebibliography}{10}
\newcommand{\enquote}[1]{``#1''}

\bibitem{bb84}
C.~H. Bennett and G.~Brassard, \enquote{{Quantum cryptography: Public key
  distribution and coin tossing},} in \emph{Proceedings of IEEE International
  Conference on Computers, Systems, and Signal Processing,}  vol. 175 (1984),
  p.~8.

\bibitem{aaronson13}
S.~Aaronson and A.~Arkhipov, \enquote{The computational complexity of linear
  optics,} {\protect\JournalTitle{Theory of Computing}} \textbf{9}, 143--252
  (2013).

\bibitem{kok07}
P.~Kok, W.~J. Munro, K.~Nemoto, T.~C. Ralph, J.~P. Dowling, and G.~J. Milburn,
  \enquote{Linear optical quantum computing with photonic qubits,}
  {\protect\JournalTitle{Rev. Mod. Phys.}} \textbf{79}, 135--174 (2007).

\bibitem{weedbrook12}
C.~Weedbrook, S.~Pirandola, R.~Garc\'{\i}a-Patr\'on, N.~J. Cerf, T.~C. Ralph,
  J.~H. Shapiro, and S.~Lloyd, \enquote{Gaussian quantum information,}
  {\protect\JournalTitle{Rev. Mod. Phys.}} \textbf{84}, 621--669 (2012).

\bibitem{wang20}
J.~Wang, F.~Sciarrino, A.~Laing, and M.~G. Thompson, \enquote{Integrated
  photonic quantum technologies,} {\protect\JournalTitle{Nature Photonics}}
  \textbf{14}, 273--284 (2020).

\bibitem{humphreys13}
P.~C. Humphreys, B.~J. Metcalf, J.~B. Spring, M.~Moore, X.-M. Jin, M.~Barbieri,
  W.~S. Kolthammer, and I.~A. Walmsley, \enquote{Linear optical quantum
  computing in a single spatial mode,} {\protect\JournalTitle{Phys. Rev.
  Lett.}} \textbf{111}, 150501 (2013).

\bibitem{humphreys14}
P.~C. Humphreys, W.~S. Kolthammer, J.~Nunn, M.~Barbieri, A.~Datta, and I.~A.
  Walmsley, \enquote{Continuous-variable quantum computing in optical
  time-frequency modes using quantum memories,} {\protect\JournalTitle{Phys.
  Rev. Lett.}} \textbf{113}, 130502 (2014).

\bibitem{lukens17}
J.~M. Lukens and P.~Lougovski, \enquote{Frequency-encoded photonic qubits for
  scalable quantum information processing,} {\protect\JournalTitle{Optica}}
  \textbf{4}, 8--16 (2017).

\bibitem{nunn13}
J.~Nunn, L.~J. Wright, C.~S\"{o}ller, L.~Zhang, I.~A. Walmsley, and B.~J.
  Smith, \enquote{Large-alphabet time-frequency entangled quantum key
  distribution by means of time-to-frequency conversion,}
  {\protect\JournalTitle{Opt. Express}} \textbf{21}, 15959--15973 (2013).

\bibitem{brecht15}
B.~Brecht, D.~V. Reddy, C.~Silberhorn, and M.~G. Raymer, \enquote{Photon
  temporal modes: A complete framework for quantum information science,}
  {\protect\JournalTitle{Phys. Rev. X}} \textbf{5}, 041017 (2015).

\bibitem{Eckstein}
A.~Eckstein, B.~Brecht, and C.~Silberhorn, \enquote{A quantum pulse gate based
  on spectrally engineered sum frequency generation,}
  {\protect\JournalTitle{Opt. Express}} \textbf{19}, 13770--13778 (2011).

\bibitem{reddy2}
D.~V. Reddy and M.~G. Raymer, \enquote{High-selectivity quantum pulse gating of
  photonic temporal modes using all-optical ramsey interferometry,}
  {\protect\JournalTitle{Optica}} \textbf{5}, 423--428 (2018).

\bibitem{sinclair14}
N.~Sinclair, E.~Saglamyurek, H.~Mallahzadeh, J.~A. Slater, M.~George,
  R.~Ricken, M.~P. Hedges, D.~Oblak, C.~Simon, W.~Sohler, and W.~Tittel,
  \enquote{Spectral multiplexing for scalable quantum photonics using an atomic
  frequency comb quantum memory and feed-forward control,}
  {\protect\JournalTitle{Phys. Rev. Lett.}} \textbf{113}, 053603 (2014).

\bibitem{ansari18}
V.~Ansari, J.~M. Donohue, B.~Brecht, and C.~Silberhorn, \enquote{Tailoring
  nonlinear processes for quantum optics with pulsed temporal-mode encodings,}
  {\protect\JournalTitle{Optica}} \textbf{5}, 534--550 (2018).

\bibitem{fan16}
L.~Fan, C.-L. Zou, M.~Poot, R.~Cheng, X.~Guo, X.~Han, and H.~X. Tang,
  \enquote{Integrated optomechanical single-photon frequency shifter,}
  {\protect\JournalTitle{Nature Photonics}} \textbf{10}, 766--770 (2016).

\bibitem{wright17}
L.~J. Wright, M.~Karpi\ifmmode~\acute{n}\else \'{n}\fi{}ski, C.~S\"oller, and
  B.~J. Smith, \enquote{Spectral shearing of quantum light pulses by
  electro-optic phase modulation,} {\protect\JournalTitle{Phys. Rev. Lett.}}
  \textbf{118}, 023601 (2017).

\bibitem{karpinski18}
F.~So\'{s}nicki and M.~Karpi\'{n}ski, \enquote{Large-scale spectral bandwidth
  compression by complex electro-optic temporal phase modulation,}
  {\protect\JournalTitle{Opt. Express}} \textbf{26}, 31307--31316 (2018).

\bibitem{TORRES}
V.~Torres~Company, J.~Lancis, and P.~Andrés, \emph{Space-Time Analogies in
  Optics} (2011), vol.~56, pp. 1--80.

\bibitem{kolner89}
B.~H. Kolner and M.~Nazarathy, \enquote{Temporal imaging with a time lens,}
  {\protect\JournalTitle{Opt. Lett.}} \textbf{14}, 630--632 (1989).

\bibitem{kolner94}
B.~H. {Kolner}, \enquote{Space-time duality and the theory of temporal
  imaging,} {\protect\JournalTitle{IEEE Journal of Quantum Electronics}}
  \textbf{30}, 1951--1963 (1994).

\bibitem{avenhaus09}
M.~Avenhaus, A.~Eckstein, P.~J. Mosley, and C.~Silberhorn,
  \enquote{Fiber-assisted single-photon spectrograph,}
  {\protect\JournalTitle{Opt. Lett.}} \textbf{34}, 2873--2875 (2009).

\bibitem{Goda}
J.~B. Goda~K., \enquote{Dispersive fourier transformation for fast continuous
  single-shot measurements,} {\protect\JournalTitle{Nature Photonics}}
  \textbf{7}, 102--112 (2013).

\bibitem{mandel-wolf}
L.~Mandel and E.~Wolf, \emph{Optical Coherence and Quantum Optics} (Cambridge
  University Press, 1995).

\bibitem{diels-rudolph}
J.-C. Diels and W.~Rudolph, eds., \emph{Ultrashort Laser Pulse Phenomena}
  (Academic Press, Burlington, 2006), second edition ed.

\bibitem{weiner11}
A.~M. Weiner, \enquote{Ultrafast optical pulse shaping: A tutorial review,}
  {\protect\JournalTitle{Optics Communications}} \textbf{284}, 3669 -- 3692
  (2011). Special Issue on Optical Pulse Shaping, Arbitrary Waveform
  Generation, and Pulse Characterization.

\bibitem{weiner}
A.~M. Weiner, \enquote{Femtosecond pulse shaping using spatial light
  modulators,} {\protect\JournalTitle{Review of Scientific Instruments}}
  \textbf{71}, 1929--1960 (2000).

\bibitem{davis17}
A.~O.~C. Davis, P.~M. Saulnier, M.~Karpi\'{n}ski, and B.~J. Smith,
  \enquote{Pulsed single-photon spectrometer by frequency-to-time mapping using
  chirped fiber bragg gratings,} {\protect\JournalTitle{Opt. Express}}
  \textbf{25}, 12804--12811 (2017).

\bibitem{Morizur}
J.-F. Morizur, L.~Nicholls, P.~Jian, S.~Armstrong, N.~Treps, B.~Hage, M.~Hsu,
  W.~Bowen, J.~Janousek, and H.-A. Bachor, \enquote{Programmable unitary
  spatial mode manipulation,} {\protect\JournalTitle{J. Opt. Soc. Am. A}}
  \textbf{27}, 2524--2531 (2010).

\bibitem{Borevich}
Z.~I. Borevich and S.~L. Krupetskii, \enquote{Subgroups of the unitary group
  that contain the group of diagonal matrices,} {\protect\JournalTitle{Journal
  of Soviet Mathematics}} \textbf{17}, 1951--1959 (1981).

\bibitem{SCHMID}
M.~Schmid, R.~Steinwandt, J.~Müller-Quade, M.~Rötteler, and T.~Beth,
  \enquote{Decomposing a matrix into circulant and diagonal factors,}
  {\protect\JournalTitle{Linear Algebra and its Applications}} \textbf{306},
  131 -- 143 (2000).

\bibitem{Wang:18}
C.~Wang, M.~Zhang, X.~Chen, M.~Bertrand, A.~Shams-Ansari, S.~Chandrasekhar,
  P.~Winzer, and M.~Lon\v{c}ar, \enquote{100-ghz low voltage integrated lithium
  niobate modulators,} in \emph{Conference on Lasers and Electro-Optics,}
  (Optical Society of America, 2018), p. SM3B.4.

\bibitem{WANG}
X.~Wang, C.-S. Yu, and X.~Yi, \enquote{An alternative quantum fidelity for
  mixed states of qudits,} {\protect\JournalTitle{Physics Letters A}}
  \textbf{373}, 58 -- 60 (2008).

\bibitem{karpinski20}
F.~Sośnicki, M.~Mikołajczyk, A.~Golestani, and M.~Karpiński,
  \enquote{Aperiodic electro-optic time lens for spectral manipulation of
  single-photon pulses,} {\protect\JournalTitle{Applied Physics Letters}}
  \textbf{116}, 234003 (2020).

\end{thebibliography}

\end{document}